# Dynamics of the CO$^+$ coma of comet 29P/Schwasmann–Wachmann 1

Oleksandra Ivanova,[1,2,3‹] Oleksiy Agapitov,[3,4‹] Dusan Odstrcil,[5,6] Pavlo Korsun,[2] Viktor Afanasiev[7] and Vera Rosenbush[2,3]

[1]*Astronomical Institute of the Slovak Academy of Sciences, SK-05960 Tatranská Lomnica*
[2]*Main Astronomical Observatory of the National Academy of Sciences of Ukraine, 27 Zabolotnoho Str., 03143 Kyiv, Ukraine*
[3]*Taras Shevchenko National University of Kyiv, 3 Observatorna Str., 04053 Kyiv, Ukraine*
[4]*Space Sciences Laboratory, University of California Berkeley, 7 Gauss Way Berkley, CA 94720, USA*
[5]*George Mason University, 4400 University Dr, Fairfax, VA 22030, USA*
[6]*NASA/GSFC, 8800 Greenbelt Rd, Greenbelt, MD 20771, USA*
[7]*Special Astrophysical Observatory of the Russian Academy of Sciences, 369167 Nizhnij Arkhyz, Russia*

**ABSTRACT**

Comet-centaur 29P/Schwassmann–Wachmann 1 was observed in CO$^+$ emission and contin-uum during 2007–2009 using the 6-m Big Telescope Alt-azimuth at the Special Astrophysical Observatory of the Russian Academy of Sciences. We analysed the morphology of the CO$^+$ and dust coma. The distributions of dust and CO$^+$ ions in the coma are not similar and vary depending on the level of comet activity. CO$^+$ ions are more concentrated towards the nucleus than the dust continuum. The column density of the CO$^+$ was derived and found to vary from $3.7 \times 10^9$ to $4.3 \times 10^{10}$ ions cm$^{-2}$. The production rate of CO$^+$ was estimated to be from $(7.01 \pm 2.7) \times 10^{24}$ to $(1.15 \pm 0.5) \times 10^{26}$ ions s$^{-1}$. We discuss possible mechanisms for the ionization of cometary material and show that impact ionization by solar wind particles is probably the main ionization mechanism at large heliocentric distances.

## INTRODUCTION

Comet 29P/Schwassmann–Wachmann 1 (hereafter, 29P) was de-tected near Jupiter in 1927 November by Arnold Schwassmann and Arno Arthur Wachmann (http://cometography.com/pcomets/029p.html). The approximately circular orbit of the comet ($q = 5.72$ au; $e = 0.044$) was formed under the influence of gravitational perturbations from Jupiter. This comet is related to the Jupiter family of comets; however, an analysis of the evolution of the comet's orbital clones (Neslusan, Tomko & Ivanova 2017) showed that it might have come to the planetary region from the Oort Cloud. Furthermore, its low orbital inclination value favours an origin in the trans-Neptunian belt. Today, this comet is often presented as a centaur (Jewitt 2009).

Comet 29P is perhaps one of the best known of the few observed distant comets, which reveal the presence of CO$^+$ ions in their coma at large distances (more than 4 au) from the Sun (Cochran, Barker & Cochran 1980; Larson 1980; Cochran, Cochran & Barker 1982; Cochran et al. 1991; Cochran & Cochran 1991; Cook, Desch & Wyckoff 2005; Korsun, Ivanova & Afanasiev 2008; Korsun et al. 2014; Ivanova et al. 2016, 2018). CO$^+$ ions have been found in the comas of comets C/1961 R1 (Humason), 29P, C/1995 O1 Hale-Bopp and C/2002 VQ94 (LINEAR) beyond Jupiter. For 29P, episodic outbursts are regularly observed (Trigo-Rodríguez et al. 2008). CO$^+$ ions have been detected both in the outbursts and in the stationary state of the comet. The slow photoionization rate of CO or CO$_2$ at heliocentric distances *greater* than 4 au cannot explain the presence of CO$^+$ in the coma of 29P.

Various physical mechanisms to explain the sources of activity in distant comets have been considered over the past decades (Wurm & Rahe 1969; Gringauz et al. 1986; Ip & Axford 1986; Kranskowsky et al. 1986; Galeev 1988; Ibadov 1988, 1993), as well as possible ionization sources, including photoionization by solar irradiation, impact ionization and the exchange of charge with the solar wind, and electric discharge in the inner coma (see Mendis & Ip 1977; Huebner & Giguere 1980; Cravens et al. 1987; Edberg et al. 2016).

At present, there is no generally accepted consensus on any one ionization mechanism, nor on the dominance of any one mechanism. However, observations indicate that a considerable proportion of cometary ions are produced in the innermost coma ('collision zone'). Owing to the low activity of such comets beyond heliocentric distances of 4 au, the solar wind flow is unimpeded by collisions with cometary neutrals and can impact the surface of the nucleus directly, causing ion-induced sputtering of surface material (Coates 1997). Because the release of cometary material from the surface by sputtering is almost proportional to the concentration of the components, this can be used to derive the composition of the nucleus. As the comet approaches the Sun, its coma becomes dense enough that the solar wind is absorbed above the surface, and the sputtering ions disappear (Nilsson et al. 2015; Wurz et al. 2015). The effects of ionization from the surface by sputtering are known from the direct observations of comet 67P/Churyumov–Gerasimenko by *Rosetta* (Wurz et al. 2015) and from observations in the vicinity of objects without any atmosphere: lunar ions sputtered directly from the surface are regularly observed in the solar wind as pickup ions (Hilchenbach et al. 1993; Halekas et al. 2012; Harada et al. 2015).

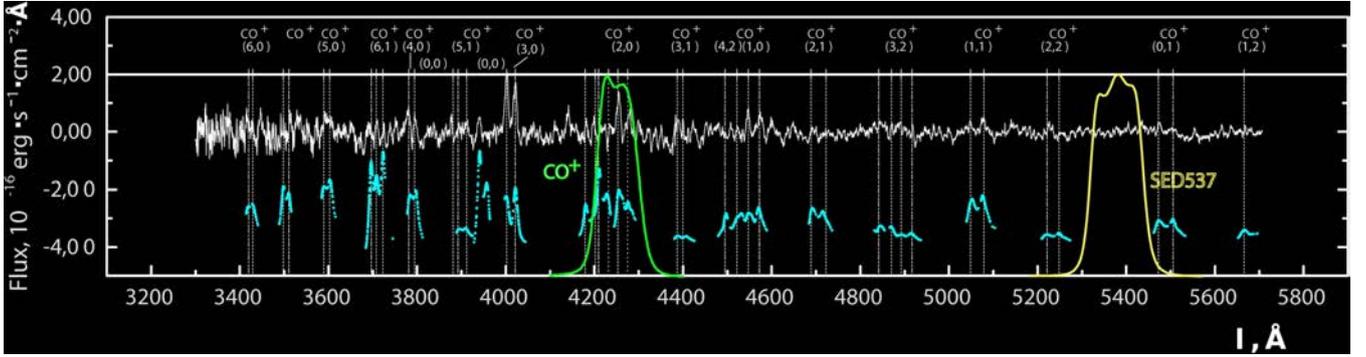

**Figure 1.** Normalized transmission curves of the CO$^+$ (λ4277/39 Å) and SED537 (λ5309/169 Å) filters superimposed on the observed spectrum of comet 29P taken from Ivanova et al. (2018). Calculated spectra of the identified species are displayed at the bottom of the figure (red denotes CO$^+$ bands).

We provide an analysis of the maps of the column density of CO$^+$ ions and the dust coma in comet 29P obtained during observations from 2007 to 2011. We also propose a physical mechanism to explain the observed ionization of CO in the coma of this comet. The paper is structured as follows. Section 2 presents the observations and data processing. Section 3 introduces a description of the solar wind model used and an analysis of the correlation of the CO$^+$ pickup ion flux with the modelled solar wind proton flux. Section 4 describes and discusses the results.

## OBSERVATIONS AND DATA REDUCTION

Comet 29P was observed photometrically using the 6-m Big Telescope Alt-azimuth (BTA) with the multimode focal reducer Spectral Camera with Optical Reducer for Photometrical and Interferometrical Observations (SCORPIO) (Afanasiev & Moiseev 2011) at the Special Astrophysical Observatory (SAO, Russia). In order to track the motion of the comet precisely, we used an offset guiding during our observations. The detector was the CCD chip (EEV 42–40) of 2048 × 2048 pixels (with a pixel scale of 0.18 arcsec pixel$^{-1}$). The matrix provides a field of view of 6.1 arcmin × 6.1 arcmin. In order to obtain a higher signal-to-noise ratio for weak cometary structures, all the obtained images were binned into 2 pixels. This procedure was carried out before reducing and analysing the data.

The comet was observed through the narrow-band CO$^+$ (λ4277/39 Å) and the middle-band filters SED537 (λ5309/169 Å). We also carried out twilight-sky observations with the same filters in order to make flat-field corrections. The images of reference stars BD+75d325, BD+33d2642, GD108 and Feige 34 obtained on different nights were used to provide photometric calibration. The transmission curves of the CO$^+$ and SED537 filters superimposed on the observed spectrum of comet 29P (Ivanova et al. 2018) are shown in Fig. 1.

The observations were processed using the software designed at the SAO RAS based on Interactive Data Language (IDL). The procedure consists of basic bias subtraction, flat-fielding, and the removal of traces of cosmic rays. We subtracted the background in each frame before stacking the images (to produce a combined image) to reduce the background spatial distribution perturbations. The sky background was estimated using the frame parts that were free from the cometary coma and background star contamination by constructing the counts histogram in the image. We selected the count corresponding to the maximum of the histogram as being the sky subtracted from the image. We aligned the resulting individual composite frames that had been sky-subtracted and flat-fielded to each other using the main brightness peaks in the coma; the peaks were obtained from the coma isophots.

We stacked the photometric images obtained in the same filter together and built the intensity maps. The seeing was conditional on atmospheric turbulence and instrumental optics and was constantly equal to about 1.7–1.9 arcsec. More details of the processing of the photometric observations are given in Korsun et al. (2008), Ivanova, Korsun & Afanasiev (2009), Ivanova et al. (2016) and Rosenbush et al. (2017).

The spectroscopic observations presented in Fig. 1 were obtained with the 1.6-m telescope of the National Laboratory for Astrophysics (Brazil) on 2011 May 31 (Ivanova et al. 2018). The instrumental setup and data processing are presented in detail in Cavichia, Costa & Maciel (2010) and Ivanova et al. (2016).

Table 1 summarizes the characteristics of our observations: the date of observation (the mid-cycle time, UT date), the geocentric (Δ) and heliocentric ($r$) distances, the phase angle of the comet ($\alpha$), the position angle of the extended Sun–comet radius vector ($\varphi$), the filter (photometry) and grid (spectroscopy), the total exposure time during the night ($T_{\rm exp}$), and the mode of the observation.

## MORPHOLOGY OF THE COMA

As can be seen in Fig. 1, the λ4277/39 Å filter transmits the *radiation* of CO$^+$ ions and the continuum, whereas the *transmitted radiation* flux through the λ5309/169 Å filter may truly be considered as the *continuum, without* CO$^+$ ions. Thus, to derive the pure ion image, we subtracted the composite image of the continuum that was obtained in the SED537 filter from the combined CO$^+$ image, controlling the particular light transmission of each filter and various continuum levels. We used values of reddening obtained from the spectral observations of comet 29P presented by Korsun et al. (2008) and Ivanova et al. (2016, 2018). The images of the comet in the two filters were centred using the central isophotes, which was closest to the maximum brightness of the comet. The panels on the left of Fig. 2 present the intensity maps of the isolated CO$^+$ coma after subtraction of the continuum intensity from the observed intensity in the CO$^+$ band, and the panels on the right present the images of the dust coma of comet 29P.

**Table 1.** Log of the observations of comet 29P/Schwassman–Wachmann 1.

| UT date | $r$ [au] | $\Delta$ [au] | $\alpha$ [°] | $\varphi$ [°] | Filter/grid | $T_{\text{exp}}$ [s] | Mode |
| --- | --- | --- | --- | --- | --- | --- | --- |
| 2007 November 8.107 | 5.96 | 5.29 | 7.4 | 266 | CO$^+$ | 1200 | Image |
| 2007 November 8.119 | 5.96 | 5.29 | 7.4 | 266 | SED537 | 600 | Image |
| 2007 November 11.049 | 5.96 | 5.26 | 7.1 | 266 | CO$^+$ | 1200 | Image |
| 2007 November 11.047 | 5.96 | 5.26 | 7.1 | 266 | SED537 | 600 | Image |
| 2007 November 18.066 | 5.96 | 5.19 | 6.2 | 263 | CO$^+$ | 2100 | Image |
| 2007 November 18.111 | 5.96 | 5.19 | 6.2 | 263 | SED537 | 600 | Image |
| 2008 December 4.021 | 6.08 | 5.40 | 7.1 | 280 | CO$^+$ | 1500 | Image |
| 2008 December 4.018 | 6.08 | 5.40 | 7.1 | 280 | SED537 | 500 | Image |
| 2009 March 29.796 | 6.11 | 5.80 | 9.1 | 100 | CO$^+$ | 1200 | Image |
| 2009 March 29.808 | 6.11 | 5.80 | 9.1 | 100 | SED537 | 600 | Image |
| 2009 December 26.987 | 6.18 | 5.53 | 7.2 | 290 | CO$^+$ | 1200 | Image |
| 2009 December 26.991 | 6.18 | 5.53 | 7.2 | 290 | SED537 | 800 | Image |
| 2011 May 31.89 | 6.25 | 6.14 | 9.3 | 112 | 300 l/mm | 2400 | Spectral |

The isolated CO$^+$ coma is clearly observed on all the dates of observation presented here (see Fig. 2): it is compact, asymmetric relative to the optocentre, and always elongated in the sunward direction, although this elongation does not coincide exactly with the direction to the Sun. The CO$^+$ coma observed on 2007 November 8 was only slightly elongated in the sunward direction, while on 2008 December 4 the comet had a very compact and strongly elongated CO$^+$ coma in the solar hemisphere to at least 60 000 km. On 2009 March 29, the comet had a small asymmetric coma with strong sunward elongation. In contrast to the CO$^+$ coma, the dust coma, formed by sunlight scattered from dust particles, was more extended and sufficiently symmetric. The cometary coma was changing until the end of 2009: the 29P activity decreased and the dust coma became more compact. The CO$^+$ coma did not change appreciably during this period; it just had a weak elongation along the sunward direction.

Because we know the column density $N$(CO$^+$) across the projected comet tail and the velocity of ion flow projected on to the sky plane, we can roughly estimate the CO$^+$ production rate $Q$(CO$^+$) in the comet, according to the technique presented in Jockers et al. (1992). In order to do this, we transformed the ion intensity image into a column density image using the g-factor given by Lutz, Womack & Wagner (1993). Table 2 presents values of the column densities of CO$^+$ in ions per centimetre squared, calculated for a cometocentric distance of ∼20 000 km. The value of the ion flow velocity is the most uncertain parameter in the estimation of the CO$^+$ production rate (Jockers et al. 1992). We calculated this parameter based on formulae derived by Jockers & Bonev (1997). The results are presented in Table 2.

## SOLAR WIND PARAMETERS

Because the effects of ionization from the surface by sputtering by the solar wind-charged particles are supposed to be important (or even to dominate) as the ionization mechanism for the comet at a heliocentric distance about 6 au, we analysed the solar wind parameters in the vicinity of the comet and compared the solar wind particle flux with the observed density of the CO$^+$ coma, taking into account that the material released from the surface by sputtering is almost proportional to the flux. Direct *in situ* measurements of the solar wind parameters in the vicinity of the cometary nucleus are not available, so we used the numerical time-dependent three-dimensional magnetohydrodynamic (MHD) model of the solar wind parameters in the heliosphere (ENLIL) provided by the NASA Community Coordinated Modelling Center (CCMC). The model is based on the numerical solution of the equations for plasma mass density, momentum and energy density, and the magnetic field in the MHD approximation (Odstrcil 2003; Odstrcil, Riley & Zhao 2004). For our purposes, it is important that the model covers ± 60° in latitude, because the comet location was out of the ecliptic plane. The model has a 360° coverage in azimuth.

As can be seen from Table 2, three values of the CO$^+$ production rate were estimated during 2007 November (the 8, 11 and 18 of November 2007), that is, during a single Carrington rotation of the Sun (the Carrington rotation number is 2063), which provides a unique opportunity to proceed the dynamics of the solar wind parameters and the corresponding CO$^+$ production changes. The snapshot of the solar wind density ($N_{\text{sun}}$) normalized to the heliocentric distance ($r$, in au) as $N_{\text{sun}} r^2$ is shown in Fig. 3(a). The period of time processed had a quiet level of Sun activity and included only one significant density enhancement, to $N_{\text{sun}} r^2 \sim 35$–40 cm$^{-3}$ (with a background level of ∼1–3 cm$^{-3}$). A comparison of the model with direct measurements at 1 au distance by the *STEREO-A*, *-B* and *Wind* spacecrafts (Fig. 3b–d) demonstrates a relatively good agreement with the averaged solar wind parameters for this enhancement (detected by all three spacecrafts) passing through the Earth on 2007 October 23–25, thus providing local verification of the model run results. This density enhancement then propagated with the average velocity of the solar wind (∼380–430 km s$^{-1}$).

The dynamics of the normalized density of the solar wind during 2007 November (from the ENLIL model calculation) in the vicinity of 29P are presented in Fig. 4. The density enhancement (localized azimuthally and radially) was passed through the comet during November 14–19 (Fig. 4a). During the calculated interval (from November 7 to November 16), the normalized density was continuously increasing. The CO$^+$ pickup ion flux exhibited a corresponding increase from November 7 to November 18, in agreement with the dynamics of the solar wind flux.

We provide the numerical calculations of the solar wind flux in the vicinity of the solar wind flux in the vicinity of 29P based on the ENLIL model runs for the dates from Table 2, and

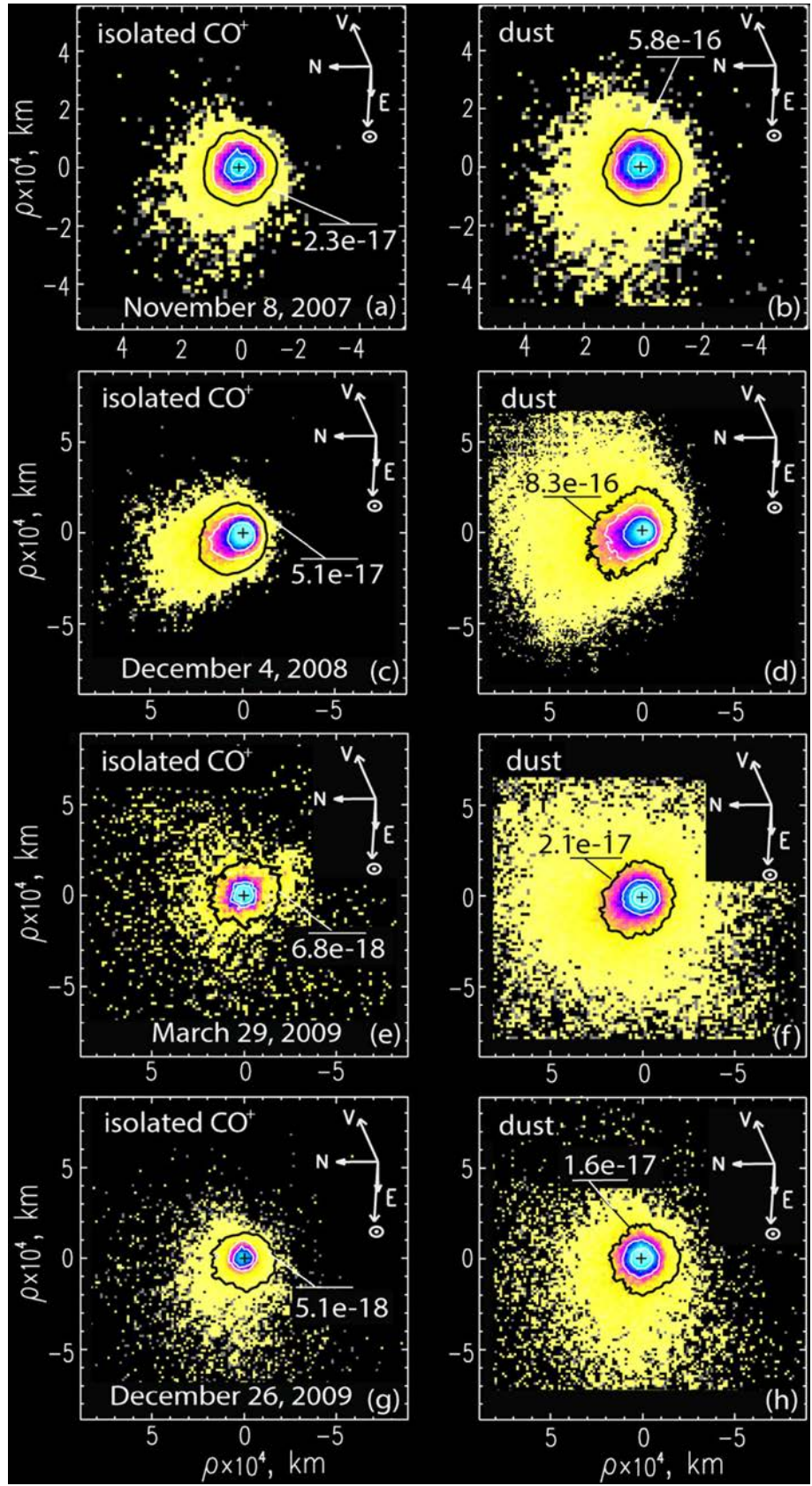

**Figure 2.** Isolated CO$^+$ (a, c, e, g) and dust (b, d, f, h) coma of comet 29P derived from observations on November 8, 2007, December 4, 2008, March 29, 2009 and December 26, 2009. All images are scaled to erg cm$^{-2}$ s$^{-1}$ Å$^{-1}$. The values near the white contours, corresponding to the column density at a cometocentric distance about 20 000 km, are presented in Table 2. The sunward (⊙), north (N) and east (E) directions, and the cometary motion vector (V) are shown.

**Table 2.** $CO^+$ production rate in comet 29P/Schwassman–Wachmann 1.

| UT date | $r$ [au][ | $\Delta$ [au] | $\alpha$ [°] | $\varphi$ [°] | $N(CO^+)$ [ions cm$^{-2}$] | $Q(CO^+)$ [ions s$^{-1}$] | Mode |
|---|---|---|---|---|---|---|---|
| 2007 November 8.107 | 5.96 | 5.29 | 7.4 | 266 | $1.60 \times 10^{10}$ | $2.49 \pm 0.7 \times 10^{25}$ | Image |
| 2007 November 11.049 | 5.96 | 5.26 | 7.1 | 266 | $2.12 \times 10^{10}$ | $3.35 \pm 0.9 \times 10^{25}$ | Image |
| 2007 November 18.066 | 5.96 | 5.19 | 6.2 | 263 | $2.84 \times 10^{10}$ | $3.94 \pm 1.1 \times 10^{25}$ | Image[a] |
| 2008 December 4.021 | 6.08 | 5.40 | 7.1 | 280 | $4.23 \times 10^{10}$ | $1.15 \pm 0.5 \times 10^{26}$ | Image |
| 2009 March 29.796 | 6.11 | 5.80 | 9.1 | 100 | $4.78 \times 10^9$ | $1.99 \pm 0.9 \times 10^{25}$ | Image |
| 2009 December 26.987 | 6.18 | 5.53 | 7.2 | 290 | $3.75 \times 10^9$ | $7.01 \pm 2.7 \times 10^{24}$ | Image |
| 2011 May 31.89 | 6.25 | 6.14 | 9.3 | 112 | $1.11 \times 10^{10}$ | $2.01 \pm 0.6 \times 10^{25}$ | Spectra |

[a]From Ivanova et al. (2009)

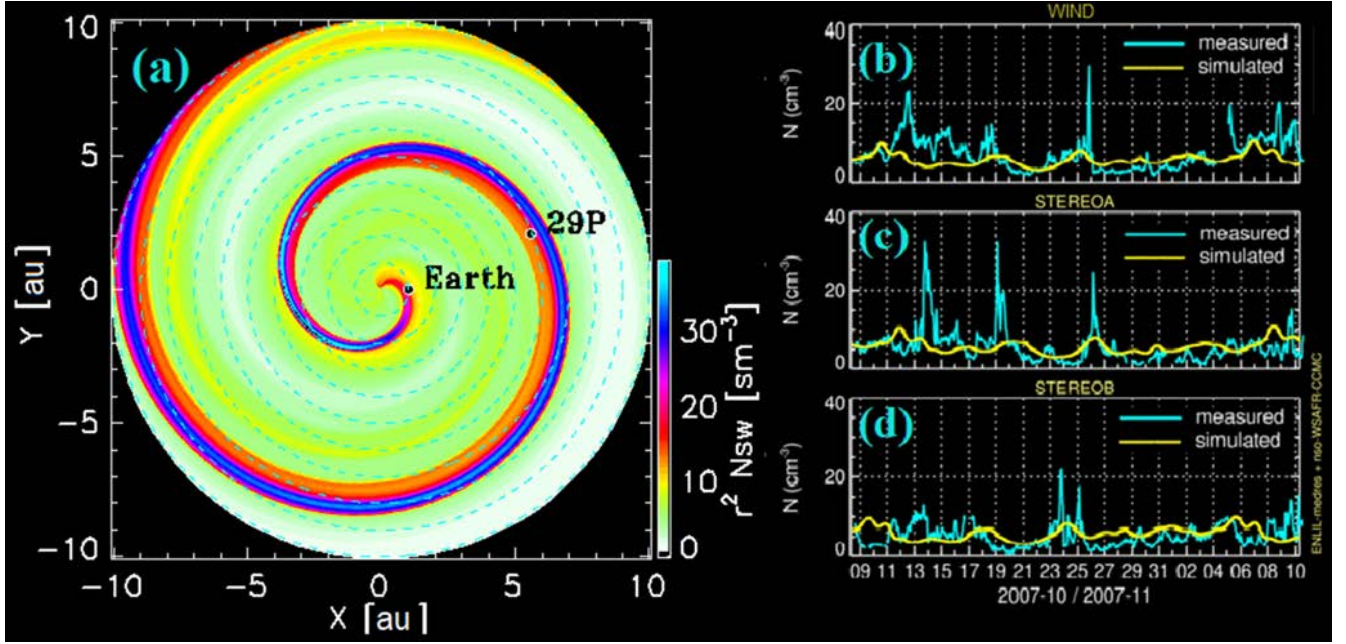

**Figure 3** (a) The normalized (multiplied by $r^2$ in au) solar wind ion density in the vicinity of the ecliptic plane on 2007 November 18 obtained from the ENLIL model. The locations of the Earth and comet 29P are indicated. Panels (b, c, d) present the solar wind density at the Earth's orbit from the plasma measurements aboard (b) WIND, (c) STEREO-A and (d) STEREO-B (red curves). The blue curves show the values obtained from the ENLIL model calculation.

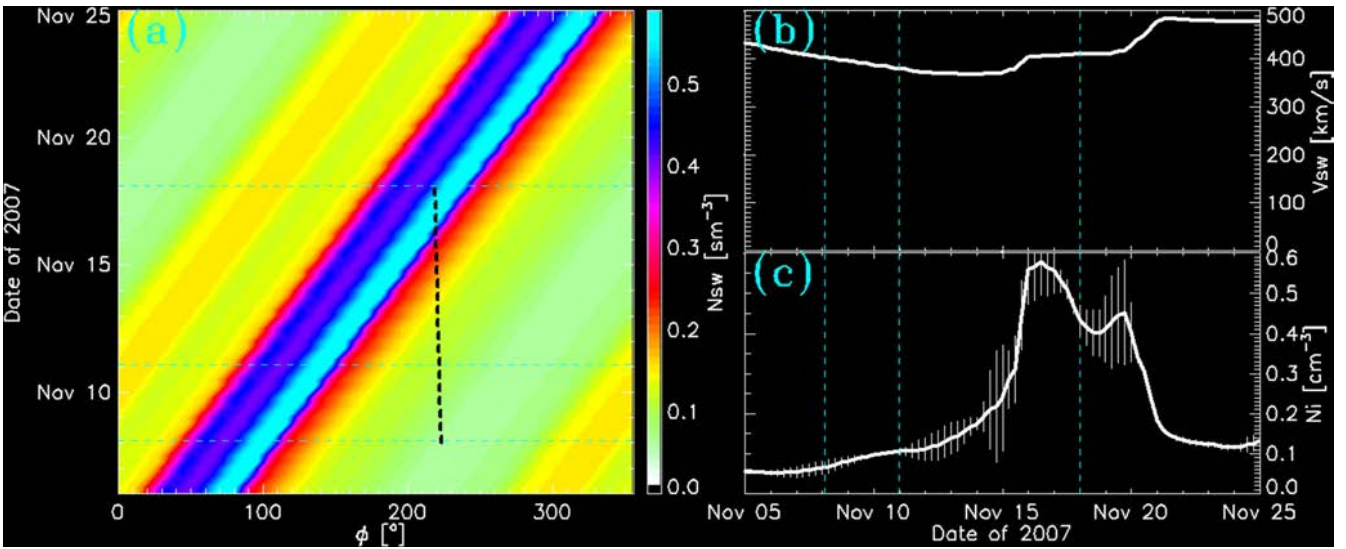

**Figure 4.** (a) The normalized solar wind flux dynamics at the comet 29P orbit. The dates of the $CO^+$ observations are indicated by the horizontal dashed lines. The 29P location is marked by the dashed white curve. Panels (b) and (c) present the dynamics of the solar wind velocity and flux in a vicinity of the comet (the profile of the comet trace from panel (a)). The dates of the CO+ observations are indicated by the red dashed lines.

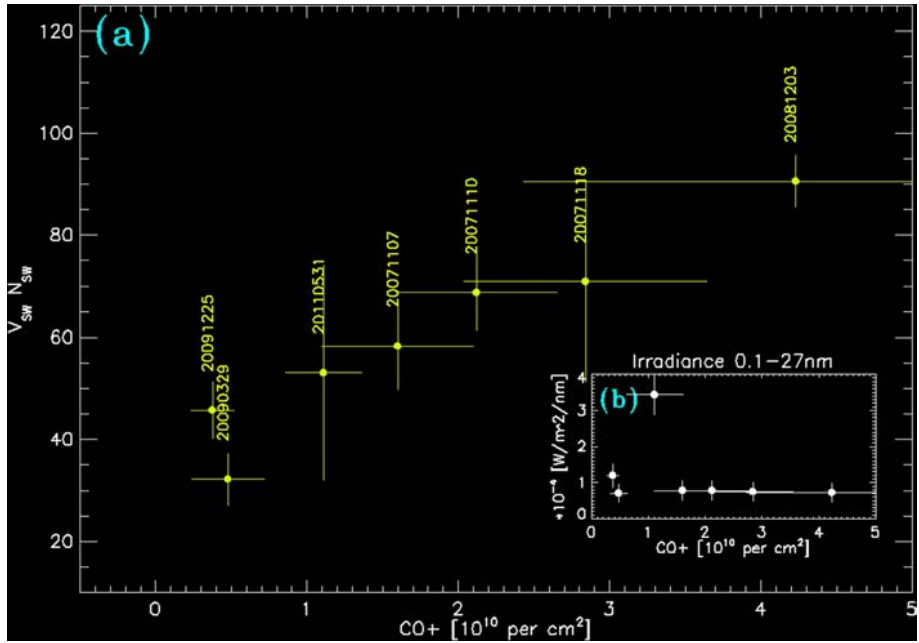

**Figure 5.** (a) The effect of the solar wind proton flux on the $CO^+$ pickup ion flux in the vicinity of comet 29P. (b) The value of the mean UV irradiance during the period of the observations.

the correspondence to the $CO^+$ production rate of the comet is presented in Fig. 5(a). The almost linear dependence indicates that the sputtering ionization, which is supposed to be proportional to the particle flux, is probably the most efficient ionization mechanism for comets at large heliocentric distances (∼6 au for 29P). Fig. 5(b) presents the daily mean UV irradiance level on the $CO^+$ production rate and shows that there is no clear dependence.

## CONCLUSIONS

Our main results can be summarized as follows:

1. $CO^+$ ions were detected in the coma of comet-centaur 29P/Schwassmann–Wachmann 1 in the set of seven observation sessions (obtained with the 6-m BTA at the SAO RAS) sampled from 2007 to 2011, and the $CO^+$ maps were obtained.

2. The column densities of $CO^+$ were derived to be from $3.75 \times 10^9$ to $4.23 \times 10^{10}$ ions cm$^{-2}$ (for the different sessions of observations).

3. We have estimated the $CO^+$ production rate to be in the range from $7.01 \times 10^{24}$ to $1.15 \times 10^{26}$ ions s$^{-1}$.

4. Processing the solar wind parameters (calculations based on the NASA CCMC ENLIL solar wind model) provided a close to linear dependence of the ionization rate on the solar wind ion flux.

5. The linear dependence of the ionization rate and the solar wind ion flux suggests that impact ionization by solar wind particles is the main ionization mechanism at large heliocentric distances.

## ACKNOWLEDGEMENTS

The observations at the 6-m BTA were performed with finan-cial support from the Ministry of Education and Science of the Russian Federation (agreement no. 14.619.21.0004, project ID RFMEFI61914 × 0004). OI is grateful for funding from the SASPRO Programme, REA grant agreement no. 609427, and the Slovak Academy of Sciences (Grant Vega no. 2/0023/18). Simulation results were provided by the Community Coordinated Modeling Center at Goddard Space Flight Center through their public Runs on Request system (http://ccmc.gsfc.nasa.gov). The ENLIL model was developed by DO at George Mason University.